\documentclass[secnumarabic, prb]{revtex4}

\begin{document}

\title{Special relativity in terms of Lie groups}
\author{Valery P.\ Dmitriyev}
\email{aether@yandex.ru}
\affiliation{Lomonosov University,  Moscow, Russia}
\date{\today}

\begin{abstract}
The special theory of relativity is constructed demanding the retention of the
 rectilinear form of a trajectory and invariance
of the wave equation under linear transformations of space and
time coordinates. The usual approach to relativity based on manipulations with the impulse of light is shown to be owing to that the symmetry of the particular solution of the wave equation coincides with the symmetry of the very wave equation. Thereof instead of the equation in partial derivatives we may deal with the algebraic form referred to as the interval.
\end{abstract}

%\keywords{special relativity, Lie groups, symmetry, mechanics}

\maketitle

\section
 {Introduction}

The central instrument of the special  relativity theory is a
pulse of light. The light is usually treated in this theory merely
as a signal with no insight into its physical nature. The only
property necessary to be specified is a peculiar feature of the
speed of delivery of the signal. In the current discourse we
construct the theory of relativity taking into account explicitly
that the light is a wave \cite{Umov}. The content of the theory
becomes the demand that the wave equation and dynamic equation of
mechanics have one and the same symmetry \cite{Zhuravlev}.

\section{Inertial frames of reference: classical definition}

Axiom. There exists at least one frame of reference $x, y, z, t$
where a free material point describes a rectilinear trajectory.
For simplicity we will consider only a one-dimensional case
\begin{equation}
 x = x_0 + u t,
\label{1}
\end{equation}
where $u$ and $x_0$ are constants. We will find other reference
frames $x', t'$ where this trajectory is rectilinear as well.
These are obviously all frames of reference which can be obtained
by affine transformations of the original reference frame. The
translation is
\begin{equation}
 x' = x + \delta,\label{2}
\end{equation}
\begin{equation}
t' = t + \tau, \label{3}
\end{equation}
where $\delta$ and $\tau$ are variable parameters. The extension is
\begin{equation}
 x' = x + \lambda x,\label{4}
\end{equation}
\begin{equation}
t' = t + \mu t. \label{5}
\end{equation}
The linear transformation that intermixes the space and time
coordinates is
\begin{equation}
 x' = x + \alpha t,\label{6}
\end{equation}
\begin{equation}
t' = t + \beta x. \label{7}
\end{equation}
The particular type of (6), (7) is given by the Galileo
transformation
\begin{equation}
 x' = x - v t,\label{8}
\end{equation}
\begin{equation}
t' = t . \label{9}
\end{equation}
All frames of reference obtained by such transformations are called
inertial frames of reference in the classical sense.

Otherwise we may define inertial frames of reference as those
which do not change the form of the equation
\begin{equation}
\frac{d^2 x}{d t^2} = 0  \label{10}
\end{equation}
that specifies the family of straight lines (\ref{1}). The form
(\ref{10}) eliminates transformations (\ref{7}) with $\beta \neq
0$ and restricts (\ref{4}), (\ref{5}) by a dissimilar extension
(see Appendix \ref{affine}).

\section{Inertial frames of reference: relativistic definition}

We introduce another restriction in the definition of inertial
frames of reference. Consider reference frames obtained by linear
transformations that do not change the equation of the
electromagnetic wave
\begin{equation}
\frac{\partial^2 \textbf{A}}{\partial t^2} = c^2 \frac{\partial^2
\textbf{A}}{\partial x^2} , \label{11}
\end{equation}
where $c$ is the speed of light. The extension (\ref{4}),
(\ref{5}) complies this requirement when it is a similarity
transformation for variables $x$ and $ct$. The transformation
(\ref{6}), (\ref{7}) does not in general leave invariant
Eq.~(\ref{11}). It works only provided that $\beta = \alpha/c^2$
\begin{equation}
 x' = x + \alpha ct,\label{12}
\label{12}
\end{equation}
\begin{equation}
 t' = t + \alpha x/c,\label{13}
\label{13}
\end{equation}
and $\alpha \rightarrow 0$. We may verify this substituting
(\ref{12}), (\ref{13}) into Eq.~(\ref{11}) and neglecting
$\alpha^2$ terms:
\begin{eqnarray}
\frac{\partial}{\partial x} &=& \frac{\partial x'}{\partial x}\frac{\partial}{\partial
x'} + \frac{\partial t'}{\partial x}\frac{\partial}{\partial
t'} = \frac{\partial}{\partial x'} +
\frac{\alpha\partial}{c\partial t'},\label{14}\\
\frac{\partial^2}{\partial x^2} &=& \frac{\partial}{\partial
x'}\left(\frac{\partial}{\partial x'} +
\frac{\alpha\partial}{c\partial t'}\right) + \frac{\alpha}{c}\frac{\partial}{\partial
t'}\left(\frac{\partial}{\partial x'} +
\frac{\alpha\partial}{c\partial t'}\right) \approx \frac{\partial^2}{\partial
x'^2}+\frac{2\alpha\partial^2}{c\partial t'\partial x'},\label{15}\\
\frac{\partial}{\partial t} &=& \frac{\partial t'}{\partial t}\frac{\partial}{\partial
t'} + \frac{\partial x'}{\partial t}\frac{\partial}{\partial
x'} = \frac{\partial}{\partial t'} +
\frac{\alpha c\partial}{\partial x'},\label{16}\\
\frac{\partial^2}{\partial t^2} &=& \frac{\partial}{\partial
t'}\left(\frac{\partial}{\partial t'} + \frac{\alpha
c\partial}{\partial x'}\right) +
\frac{\alpha c\partial}{\partial x'}\left(\frac{\partial}{\partial t'} +
\frac{\alpha c\partial}{\partial x'}\right)\approx \frac{\partial^2}{\partial
t'^2}+\frac{2\alpha c\partial^2}{\partial t'\partial x'}.\label{17}
\end{eqnarray}

\section{Lie groups}

We may try to construct a finite transformation by a successive
application of infinitesimal steps (\ref{12}), (\ref{13}):
\begin{equation}
x'' = x' + \beta c t' = x + \alpha c t + \beta c (t + \alpha x/c)
\approx x + \gamma c t, \label{18}
\end{equation}
\begin{equation}
t'' = t' + \beta x'/ c  = t + \alpha x/c + \beta  (x + \alpha
ct)/c \approx t + \gamma x/c , \label{19}
\end{equation}
where
\begin{equation}
\gamma = \alpha + \beta . \label{20}
\end{equation}
Relations (\ref{12}), (\ref{13}) and (\ref{18}), (\ref{19}) with (\ref{20}) say that these infinitesimal transformations form a one-parameter Lie group.

In general, transformations
\begin{equation}
x' = \Phi(x, t, \alpha), \quad t' = \Psi(x, t, \alpha) \label{21}
\end{equation}
form a group with the parameter $\alpha$ if from (\ref{21}) and
\begin{equation}
x'' = \Phi(x', t', \beta), \quad t'' = \Psi(x', t', \beta)
\label{22}
\end{equation}
follows
\begin{equation}
x'' = \Phi(x, t, \gamma), \quad t'' = \Psi(x, t, \gamma)
\label{23}
\end{equation}
with the group operation
\begin{equation}
\gamma = \varphi(\alpha,\beta) . \label{24}
\end{equation}
For example, similarity transformations
\begin{equation}
x' = x + \alpha x, \quad t' =  t + \alpha t \label{25}
\end{equation}
form a Lie group with the group operation
\begin{equation}
\gamma = \alpha + \beta + \alpha \beta. \label{26}
\end{equation}
The group parameter is said to be canonical if it is additive as
in (\ref{20}).

The coordinates transformation group is the way that the transition between certain  references frames can be parameterized.

We can always accommodate the group parameter $\alpha$ so that  the identity transformation will correspond to $\alpha = 0$. Then, transformation (\ref{21}) can be expanded into the Taylor series with the linear part
\begin{equation}
x' = x + \alpha \zeta(x,t) + ... , \quad t' =  t + \alpha \eta(x, t) + ... ,
\label{27}
\end{equation}
where functions $\zeta(x,t)$ and $\eta(x, t)$ are referred to as
the kernel of the group. It can be shown (the second Lie theorem)
that the kernel $\zeta, \eta$ of the group having been given we
may restore the whole transformation $\Phi, \Psi$ solving the
following set of ordinary differential equations; when the group
parameter $\alpha$ is canonical these equations are
\begin{equation}
\frac{dx'}{d\alpha} =\zeta(x',t'), \quad \frac{dt'}{d\alpha} =
\eta(x', t') \label{28}
\end{equation}
(see Appendix \ref{lie}).

\section{The Lorentz group}

Applying Eqs.~(\ref{27}), (\ref{28}) to (\ref{12}), (\ref{13})
with the account of (\ref{20}) we can write down the following
equations
\begin{equation}
\frac{dx'}{d\alpha} =ct', \quad \frac{dt'}{d\alpha} = x'/c.
\label{29}
\end{equation}
Eqs.~(\ref{29}) have the solution
\begin{equation}
x' = x\cosh\alpha + ct\sinh\alpha, \quad ct' = x\sinh\alpha +
ct\cosh\alpha . \label{30}
\end{equation}
Substituting (\ref{30}) in
\begin{equation}
x'' = x'\cosh\beta + ct'\sinh\beta, \quad ct'' = x'\sinh\beta +
ct'\cosh\beta , \label{31}
\end{equation}
we may verify that the transformation (\ref{30}) is a group and
the group parameter is canonical:
\begin{eqnarray}
x'' &=& (x\cosh\alpha + ct\sinh\alpha)\cosh\beta + (x\sinh\alpha +
ct\cosh\alpha)\sinh\beta \label{32}\\
 &=& x(\cosh\alpha\cosh\beta +
\sinh\alpha\sinh\beta) + ct(\sinh\alpha\cosh\beta +
\cosh\alpha\sinh\beta)\nonumber\\
 &=& x\cosh(\alpha+\beta) +
ct\sinh(\alpha+\beta),\nonumber
\end{eqnarray}
\begin{eqnarray}
ct'' &=& (x\cosh\alpha + ct\sinh\alpha)\sinh\beta + (x\sinh\alpha
+ ct\cosh\alpha)\cosh\beta \label{33}\\
&=& x(\cosh\alpha\sinh\beta + \sinh\alpha\cosh\beta) +
ct(\sinh\alpha\sinh\beta + \cosh\alpha\cosh\beta)\nonumber\\
&=& x\sinh(\alpha+\beta) + ct\cosh(\alpha+\beta).\nonumber
\end{eqnarray}
Transforming partial derivatives as in (\ref{14})-(\ref{17}) we may be convinced another time that (\ref{30}) represents a symmetry transformation of the equation (\ref{11}) (see Appendix {\ref{symmetry}).

Belonging of coordinates transformation to a symmetry group indicates that all reference frames defined by the transformation enjoy equal rights in relation to the property admitting this symmetry group. In other words  in the bounds of the Lorentz group, i.e. among inertial frames of reference, there is no a preferable reference frame.

\section{An extension to derivatives}

We may attach a physical sense to the parameter of the Lorentz
group (\ref{30}). To this end we will construct the corresponding infinitesimal
group transformation for the velocity
\begin{equation}
\dot x' = \frac{dx'}{dt'} .\label{34}
\end{equation}
Substituting (\ref{12}) and (\ref{13}) in (\ref{34}) and
neglecting terms with $\alpha^2$ we obtain
\begin{equation}
\dot x' = \frac{dx'}{dt'} = \frac{dx+\alpha cdt}{dt+\alpha dx/c}=
\frac{\dot x + \alpha c}{1 + \alpha \dot x/c} \approx \dot x +
(c-\dot x^2/c)\alpha . \label{35}
\end{equation}
From (\ref{35}) we may construct the differential equation for the
respective group transformation
\begin{equation}
\frac{d\dot x'}{d\alpha} = c- \dot x'^2/c . \label{36}
\end{equation}
The solution to (\ref{36}) is given by
\begin{equation}
\ln\left|\frac{1+\dot x'/c}{1+\dot x/c}\right| -
\ln\left|\frac{1-\dot x'/c}{1-\dot x/c}\right| = 2\alpha.
\label{37}
\end{equation}

Let the frame of reference $x', t'$ moves with the velocity $v$
relative to the fixed frame of reference $x, t$. Then we have for
the origin of the  reference frame $x', t'$:  $\dot x' = 0$ and
$\dot x = v$. Substituting this to (\ref{37}) we find
\begin{equation}
\alpha = \frac{1}{2}\ln\left|\frac{1-v/c}{1+v/c}\right|.
\label{38}
\end{equation}
Substituting (\ref{38}) in (\ref{30}) gives the Lorentz
transformation of space and time coordinates
\begin{equation}
x' = \frac{x-vt}{\sqrt{1-v^2/c^2}}, \quad t' =
\frac{t-xv/c^2}{\sqrt{1-v^2/c^2}}. \label{39}
\end{equation}
Substituting (\ref{38}) in (\ref{37}) gives the respective group
transformation of the velocity $\dot x$
\begin{equation}
\dot x' = \frac{\dot x-v}{1 - \dot x v/c^2}. \label{40}
\end{equation}

\section{Relativistic mechanics}

Now we must correct Eq.~(\ref{10}) in order it will be invariant
under the Lorentz transformation (\ref{39}). To this end we will find
the infinitesimal group transformation for the acceleration
\begin{equation}
\ddot x' = \frac{d\dot x'}{dt'}. \label{41}
\end{equation}
Substituting (\ref{35}) and (\ref{13}) in (\ref{41}) and
neglecting terms with $\alpha^2$ we obtain
\begin{equation}
\ddot x' = \frac{d\dot x'}{dt'} = \frac{d\dot x - \alpha 2\dot x d
\dot x/c}{dt+\alpha dx/c}= \frac{\ddot x - \alpha 2\dot x\ddot
x/c}{1+\alpha\dot x/c} \approx \ddot x - 3\dot x\ddot x\alpha /c .
\label{42}
\end{equation}
From (\ref{42}) we can find the differential equation for the
respective group transformation
\begin{equation}
\frac{d\ddot x'}{d\alpha} =  - 3\dot x'\ddot x'/c. \label{43}
\end{equation}

In general, we are searching the form $G(x', t', \dot x', \ddot
x')$ that does not change under the extended Lorentz
transformation, i.e.
\begin{equation}
\frac{d G}{d\alpha} = 0. \label{44}
\end{equation}
Using (\ref{29}), (\ref{36}) and (\ref{43}) we find
\begin{eqnarray}
\frac{d G}{d\alpha} &=& \frac{\partial G}{\partial
t'}\frac{d t' }{d\alpha} + \frac{\partial
G}{\partial x'}\frac{d x' }{d\alpha} +
\frac{\partial G}{\partial \dot x'}\frac{d\dot x'
}{d\alpha} + \frac{\partial G}{\partial \ddot
x'}\frac{d\ddot x'
}{d\alpha} \nonumber \\
&=& \frac{x'}{c}\frac{\partial G}{\partial t'} + ct'\frac{\partial
G }{\partial x'} + (c-\frac{\dot x'^2}{c})\frac{\partial
G}{\partial \dot x'} - \frac{3\dot x'\ddot x'}{c} \frac{\partial
G}{\partial \ddot x'} . \label{45}
\end{eqnarray}
Insofar as the form sought for does not depend on $x$ and $t$ we
must find the solution $G_3(\dot x', \ddot x')$ to the following
equation in partial derivatives
\begin{equation}
(c-\frac{\dot x'^2}{c})\frac{\partial G_3}{\partial \dot x'} -
\frac{3\dot x'\ddot x'}{c} \frac{\partial G_3}{\partial \ddot x'}
= 0. \label{46}
\end{equation}
The differential invariant $G_3$ can be found as the integral of
the respective ordinary differential equation constructed from
(\ref{46})
\begin{equation}
\frac{d \dot x'}{c-\dot x'^2/c} =  - \frac{c d \ddot x'}{3\dot
x'\ddot x'}. \label{47}
\end{equation}
This integral is
\begin{equation}
G_3 = \frac{ \ddot x'}{(1-\dot x'^2/c^2)^{3/2}} . \label{48}
\end{equation}
The form (\ref{48}) should replace the left-hand part of
Eq.~(\ref{10}):
\begin{equation}
\frac{ \ddot x}{(1-\dot x^2/c^2)^{3/2}} = 0. \label{49}
\end{equation}

Equation (\ref{49}) defines a rectilinear trajectory  (\ref{1}).
Eqs.~(\ref{49}) and (\ref{11}) are invariant under the Lorentz
transformation (\ref{30}) or (\ref{39}) extended according to
Eqs.~(\ref{36}) and (\ref{43}). Because of the difference in
symmetries of equations (\ref{49}) and (\ref{11}), extensions are
excluded from the consideration. Thus, the class of inertial
reference frames is defined by the Lorentz group, and the space
and time translations.

Using (\ref{48}) we may construct the relativistic form of the
second law of classical mechanics
\begin{equation}
\frac{d}{dt}\left[\frac{m \dot x}{(1-\dot x^2/c^2)^{1/2}}\right] =
F. \label{50}
\end{equation}

\section{Interval}

Special relativity is usually constructed \cite{Landau} starting from the notion of the interval
\begin{equation}
x^2 - c^2t^2 . \label{51}
\end{equation}
Interval (\ref{51}) is invariant under the Lorentz transformation (\ref{39}). In this section we will establish a relation between the standard approach and the approach developed in the current report.

First of all we will notice that the form (\ref{51}) can be obtained as a differential invariant $G_1$ from the first two terms of (\ref{45})
\begin{equation}
\frac{x}{c}\frac{\partial G}{\partial t} + ct\frac{\partial
G }{\partial x} = 0. \label{52}
\end{equation}
Therefore when studying the symmetry of vacuum we may deal with the algebraic expression (\ref{51}) instead of the equation (\ref{11}) in partial derivatives.

Secondly I shall explain why in the phenomenological theory of relativity it is sufficient to use properties of the light signal. The general solution of the d'Alembert equation (\ref{11}) is
\begin{equation}
\Phi_1(x-ct) + \Phi_2(x+ct) \label{53}
\end{equation}
where $\Phi_1$ and $\Phi_2$ are arbitrary functions. A differential equation having a symmetry does not imply that a solution of this equation also possesses the same symmetry. However we may find a particular solution of the wave equation whose symmetry coincides with the symmetry of the very wave equation. Considering a point disturbance emitted from the origin of coordinates we have for (\ref{53})
\begin{equation}
\delta(x-ct) + \delta(x+ct). \label{54}
\end{equation}
The following relation follows from general properties of the $\delta$-function
\begin{equation}
\delta(x-ct) + \delta(x+ct) = 2a\delta(x^2-c^2t^2) \label{55}
\end{equation}
where $a>0$ is a constant. Formula (\ref{55}) can be verified, say, integrating it over $x$ and taking $a = ct$ (see Appendix \ref{wave}). The form in the right-hand part of (\ref{55}) is invariant under the Lorentz transformation (\ref{39}) since any function of the invariant is the invariant of the group. Physically the form (\ref{54}) corresponds to a light impulse. Thus, the special theory of relativity can be constructed dealing only with a light signal.

\appendix
\section{Invariance of classical mechanics}
\label{affine}
We have for the affine transformation
\begin{eqnarray}
x' &=& \Lambda x + \alpha t + \delta , \label{a1}\\
t' &=& \beta x + \Omega t + \tau . \label{a2}
\end{eqnarray}
Taking the differential of (\ref{a1}) and (\ref{a2}):
\begin{eqnarray}
dx' &=& \Lambda dx + \alpha dt , \label{a3}\\
dt' &=& \beta dx + \Omega dt . \label{a4}
\end{eqnarray}
Dividing (\ref{a3}) by (\ref{a4}):
\begin{equation}
\frac{dx'}{dt'} = \frac{\Lambda dx + \alpha dt}{\beta dx + \Omega dt} = \frac{\Lambda\frac{dx}{dt} + \alpha}{\beta\frac{dx}{dt} + \Omega}. \label{a5}
\end{equation}
Rewriting (\ref{a5}) into the linear form:
\begin{equation}
\frac{dx'}{dt'}(\beta\frac{dx}{dt} + \Omega) = \Lambda\frac{dx}{dt} + \alpha. \label{a6}
\end{equation}
Differentiating (\ref{a6}):
\begin{equation}
(\beta\frac{dx}{dt} + \Omega)d\frac{dx'}{dt'} + \beta\frac{dx'}{dt'}d\frac{dx}{dt} = \Lambda d\frac{dx}{dt}. \label{a7}
\end{equation}
Dividing (\ref{a7}) by (\ref{a4}) we get
\begin{equation}
\frac{d^2x'}{dt'^2} = \frac{d^2x}{dt^2}\frac{\Lambda - \beta\frac{dx'}{dt'}}{(\Omega + \beta\frac{dx}{dt})^2}. \label{a8}
\end{equation}
The retention of the form (\ref{10}) requires
\begin{equation}
\beta = 0,\quad\Lambda = \Omega^2. \label{a9}
\end{equation}
Substituting (\ref{a9}) into (\ref{a1}), (\ref{a2}) we obtain the affine symmetry transformation of the form (\ref{10}):
\begin{eqnarray}
x' &=& \Omega^2 x + \alpha t + \delta , \label{a10}\\
t' &=& \Omega t + \tau . \label{a11}
\end{eqnarray}

\section{The second Lie theorem}
\label{lie}

We consider the transformation of the variable $x$ to $x'$
\begin{equation}
x' = T(x, \alpha) \label{b1}
\end{equation}
which depends on the parameter $\alpha$ so that
\begin{equation}
x = T(x, 0). \label{b2}
\end{equation}
Further, let $x'$ be transformed to $x''$:
\begin{equation}
x'' = T(x',\beta). \label{b3}
\end{equation}
Substituting (\ref{b1}) into the right-hand part of (\ref{b3}) we obtain the composition of the two transformations:
\begin{equation}
x'' = T(T(x,\alpha),\beta). \label{b4}
\end{equation}
We are interested in the case when (\ref{b4}) belongs to the same set of transformations as (\ref{b1}) and (\ref{b3}):
\begin{equation}
x'' = T(x,\gamma) \label{b5}
\end{equation}
where
\begin{equation}
\gamma = \varphi(\alpha,\beta). \label{b6}
\end{equation}
If (\ref{b1}) and (\ref{b3}) imply (\ref{b5}), (\ref{b6}) then we say that (\ref{b1}), (\ref{b3}) and (\ref{b5}) form a Lie group with the group operation (\ref{b6}).

The function (\ref{b1}) can be expanded into the Taylor series. We have with the account of (\ref{b2})
\begin{equation}
T(x,\alpha) = x + \frac{\alpha}{1!}[\partial_{\alpha}T(x,\alpha)]_{\alpha=0} +  \frac{\alpha^2}{2!}[\partial_{\alpha}^2T(x,\alpha)]_{\alpha=0} +  \frac{\alpha^3}{3!}[\partial_{\alpha}^3T(x,\alpha)]_{\alpha=0} + ... . \label{b7}
\end{equation}
So, in order to define a function $T(\alpha)$ we must know all its derivatives for $\alpha=0$.

\textbf{Theorem.}
The group $T(x,\alpha)$ is fully defined by its first derivative $\partial_{\gamma}T(x,\gamma)$ at $\gamma=0$.

\textit{Proof.}
We will make successively the following transitions
\begin{eqnarray}
&\alpha^{-1}& \quad \alpha\nonumber\\
x'&\rightarrow&x \rightarrow x' . \label{b8}
\end{eqnarray}
where $\alpha^{-1}$ is the parameter of the inverse transformation ($\alpha^{-1}\neq 1/\alpha $). The first transition in (\ref{b8}) is realized by
\begin{equation}
x = T(x',\alpha^{-1}). \label{b9}
\end{equation}
Substituting (\ref{b9}) in (\ref{b1}) we obtain similarly to (\ref{b4}), (\ref{b5})
\begin{equation}
x' = T(T(x',\alpha^{-1}),\alpha) = T(x',\gamma) \label{b10}
\end{equation}
where by (\ref{b6}) and with the account of (\ref{b2})
\begin{equation}
\gamma = \varphi(\alpha^{-1},\alpha) = 0. \label{b11}
\end{equation}
Differentiating (\ref{b10}) with respect to $\alpha$ we obtain with the account of (\ref{b11})
\begin{equation}
\frac{dx'}{d\alpha} = [\partial_{\gamma}T(x',\gamma)]_{\gamma=0}\partial_{\alpha}\varphi(\alpha^{-1},\alpha). \label{b12}
\end{equation}
The autonomous ordinary differential equation (\ref{b12}) defines unambiguously the group of the transformations (\ref{b1}).

\section{Symmetry transformation of the wave equation}
\label{symmetry}
We transform partial derivatives $\partial^2/\partial x^2$ and $\partial^2/\partial t^2$ to $\partial^2/\partial x'^2$ and $\partial^2/\partial t'^2$, respectively, by (\ref{30})
\begin{eqnarray}
x' &=& x\cosh\alpha + ct\sinh\alpha, \label{c1}\\
ct' &=& x\sinh\alpha + ct\cosh\alpha . \label{c2}
\end{eqnarray}
Transforming first derivatives with (\ref{c1}) and (\ref{c2}):
\begin{eqnarray}
\frac{\partial}{\partial x} &=& \frac{\partial x'}{\partial x}\frac{\partial}{\partial x'} + \frac{\partial t'}{\partial x}\frac{\partial}{\partial t'} = \cosh\alpha\frac{\partial}{\partial x'} + \frac{1}{c}\sinh\alpha\frac{\partial}{\partial t'}, \label{c3}\\
\frac{\partial}{\partial t} &=& \frac{\partial x'}{\partial t}\frac{\partial}{\partial x'} + \frac{\partial t'}{\partial t}\frac{\partial}{\partial t'} = c\sinh\alpha\frac{\partial}{\partial x'} + \cosh\alpha\frac{\partial}{\partial t'} . \label{c4}
\end{eqnarray}
Calculating second derivatives with (\ref{c3}) and (\ref{c4}):
\begin{eqnarray}
\frac{\partial^2}{\partial x^2} &=& (\cosh\alpha\frac{\partial}{\partial x'} + \frac{1}{c}\sinh\alpha\frac{\partial}{\partial t'})(\cosh\alpha\frac{\partial}{\partial x'} + \frac{1}{c}\sinh\alpha\frac{\partial}{\partial t'})\nonumber\\ &=& \cosh^2\alpha\frac{\partial^2}{\partial x'^2} + \frac{2}{c}\sinh\alpha\cosh\alpha\frac{\partial^2}{\partial x'\partial t'} + \frac{1}{c^2}\sinh^2\alpha\frac{\partial^2}{\partial t'^2}, \label{c5}\\
\frac{\partial^2}{\partial t^2} &=& (c\sinh\alpha\frac{\partial}{\partial x'} + \cosh\alpha\frac{\partial}{\partial t'})(c\sinh\alpha\frac{\partial}{\partial x'} + \cosh\alpha\frac{\partial}{\partial t'})\nonumber\\ &=& c^2\sinh^2\alpha\frac{\partial^2}{\partial x'^2} + 2c\sinh\alpha\cosh\alpha\frac{\partial^2}{\partial x'\partial t'} + \cosh^2\alpha\frac{\partial^2}{\partial t'^2}. \label{c6}
\end{eqnarray}
Substituting (\ref{c5}) and (\ref{c6}) into (\ref{11}) and using the identity $\cosh^2\alpha - \sinh^2\alpha = 1$ we obtain the same form of the wave equation:
\begin{equation}
\frac{\partial^2 \textbf{A}}{\partial t'^2} = c^2 \frac{\partial^2
\textbf{A}}{\partial x'^2} . \label{c7}
\end{equation}

\section{The wave front}
\label{wave}
Integrating the left-hand part of (\ref{55}):
\begin{equation}
\int\limits_{-\infty}^\infty\delta(x-ct)dx + \int\limits_{-\infty}^\infty\delta(x+ct)dx = 2. \label{d1}
\end{equation}
Integrating the right-hand part of (\ref{55}):
\begin{equation}
2a\int\limits_{-\infty}^\infty\delta(x^2-c^2t^2)dx =  4a\int\limits_0^\infty\delta(x^2-c^2t^2)dx = 2a\int\limits_0^\infty\frac{1}{x}\delta(x^2-c^2t^2)dx^2 = \frac{2a}{ct}. \label{d2}
\end{equation}
Taking in (\ref{d2}) $a=ct$ we obtain (\ref{d1}).

\end{document}